\documentclass[12pt]{iopart}
\usepackage{iopams}
\usepackage{epsfig}
\usepackage{caption2}

\def\bea{\begin{eqnarray}}  \def\eea{\end{eqnarray}}
\def\beq{\begin{equation}}   \def\eeq{\end{equation}}

\def\beeq{\begin{eqnarray}} \def\eeeq{\end{eqnarray}}

\newenvironment{narrow}[2]{%
\begin{list}{}{%
\setlength{\topsep}{0pt}%
\setlength{\leftmargin}{#1}%
\setlength{\rightmargin}{#2}%
\setlength{\listparindent}{\parindent}%
\setlength{\itemindent}{\parindent}%
\setlength{\parsep}{\parskip}}%
\item[]}{\end{list}}

\begin{document}

\title {Pion spectra and HBT radii at RHIC and LHC}

\author{Yu.M. Sinyukov, S.V. Akkelin, Iu.A. Karpenko }

\address{BITP, Metrolohichna str. 14-b, 03680 Kiev-143, Ukraine}
\begin{abstract}
We describe RHIC pion data in central A+A collisions and make
predictions for LHC based on hydro-kinetic model, describing
continuous 4D particle emission, and initial conditions taken from
Color Glass Condensate (CGC) model.
\end{abstract}

Hydro-kinetic approach to heavy ion collisions proposed in Ref.
\cite{AkkSin1} accounts for  continuous particle emission from 4D
volume of hydrodynamically expanding fireball as well as back
reaction of the emission  on the fluid dynamics. The approach is
based on the generalized relaxation time approximation  for
relativistic finite expanding systems,
\begin{eqnarray}
\frac{p^{\mu }}{p_{0}}\frac{\partial f(x,p)}{\partial x^{\mu }} =
- \frac{f(x,p)-f^{l.eq.}(x,p)}{\tau_{rel}(x,p)}, \label{boltz-1}
\end{eqnarray}
where $f(x,p)$ is phase-space distribution function (DF),
$f^{(l.eq.)}(x,p)$ is  local equilibrium distribution and
$\tau_{rel}(x,p)$ is relaxation time, $\tau_{rel} (x,p)$ as well
as $f^{l.eq}$ are functional of hydrodynamic variables. Complete
algorithm described in detail in Ref. \cite{AkkSin2} includes:
solution of equations of ideal hydro; calculation of a non local
equilibrium DF and emission function  in the first approximation;
solution of equations for ideal hydro with non-zero
right-hand-side that accounts for conservation laws at the
particle emission  during expansion; calculation of "improved" DF
and emission function; evaluation  of spectra and Bose-Einstein
correlations. Here we present our results for the pion momentum
spectra and interferometry  radii calculated for RHIC and LHC
energies in the first approximation of the hydro-kinetic approach.

 For simulations we utilize ideal fluid  model
\cite{Hirano} and realistic equation of state (EoS) that combines
high temperature  EoS with crossover transition \cite{Laine} adjusted to the QCD lattice
data and EoS of hadron resonance gas with partial chemical
equilibrium \cite{Hirano}. The gradual disappearance of pions
during the crossover transition to deconfinement and different
intensity of interactions of pions in pure hadronic and "mixed"
phases are taken into account in the  hydro-kinetic model (HKM),
but resonance contribution to pion spectra and interferometry
radii is not taken into account in the present  version of the HKM. We
assume the following initial conditions at  proper time
$\tau_{0}=1$ fm/c for HKM calculations: boost-invariance of a
system in longitudinal direction and cylindrical symmetry with
Gaussian energy density profile in transverse plane. The maximal
energy densities at RHIC, $\epsilon_{0}= 30$ GeV/fm$^3$ and at
LHC, $\epsilon_{0}=70$ GeV/fm$^3$, were calculated from Ref.
\cite{Lappi} in approximation of Bjorken expansion of free
ultrarelativistic partons till $\tau_{0}$ and adjusted for
transverse Gaussian density profile. The (pre-equilibrium) initial
transverse flows at $\tau_{0}$ were estimated assuming again a
free-streaming of partons, with transverse modes distributed
according to CGC picture, from proper time $\approx$ 0.1 fm/c till
$\tau_{0}=1$ fm/c. Finally, we approximate the transverse velocity
profile by $ v_T=\tanh (\alpha \cdot{r_T\over R_T})$ where $\alpha
=0.2$ both for RHIC and LHC energies and we suppose the fitting
Gaussian radius for RHIC top energy, $R_{T}=4.3$, to be the same
for LHC energy. Our results for RHIC and predictions for LHC are
presented in Fig. $1$. The relatively small increase of the
interferometry  radii with energy in HKM calculations  is
determined by early (as compare to sharp freeze-out prescription)
emission of hadrons, and also by increase of transverse flow at
LHC caused by longer time of expansion. It is noteworthy that in
the case of EoS related to first order phase transition, 
the satisfactory fitting of the RHIC
HBT data requires non-realistic high initial
transverse flows at $\tau_{0}=1$ fm/c: $\alpha =0.3$.

\begin{figure}
\begin{narrow}{-0.0in}{-0.0in}
\centering
\begin{minipage}[c]{1.0\textwidth}
 \includegraphics[width=3.2in]{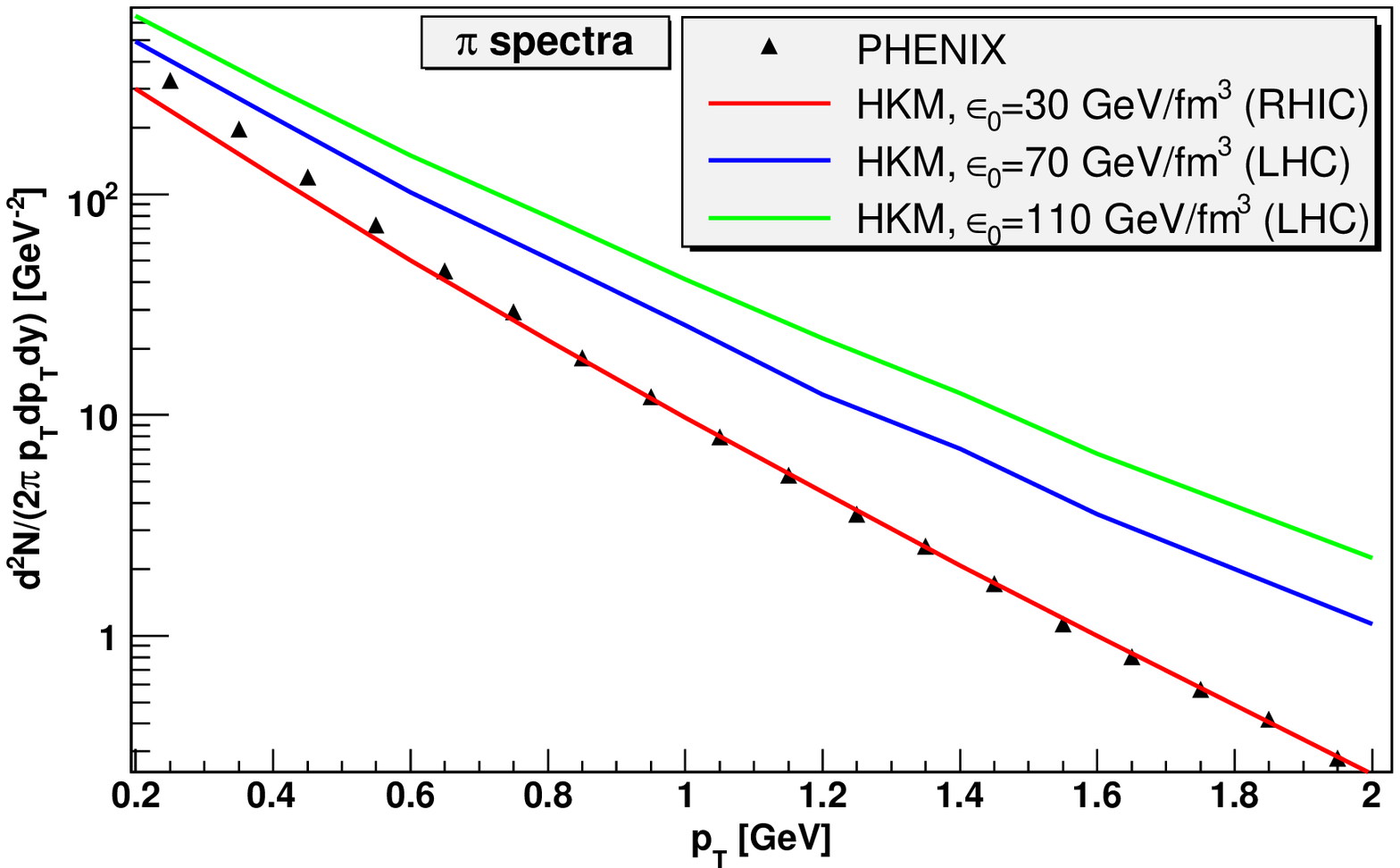}
\hspace{-0.3in}
\includegraphics[width=3.2in]{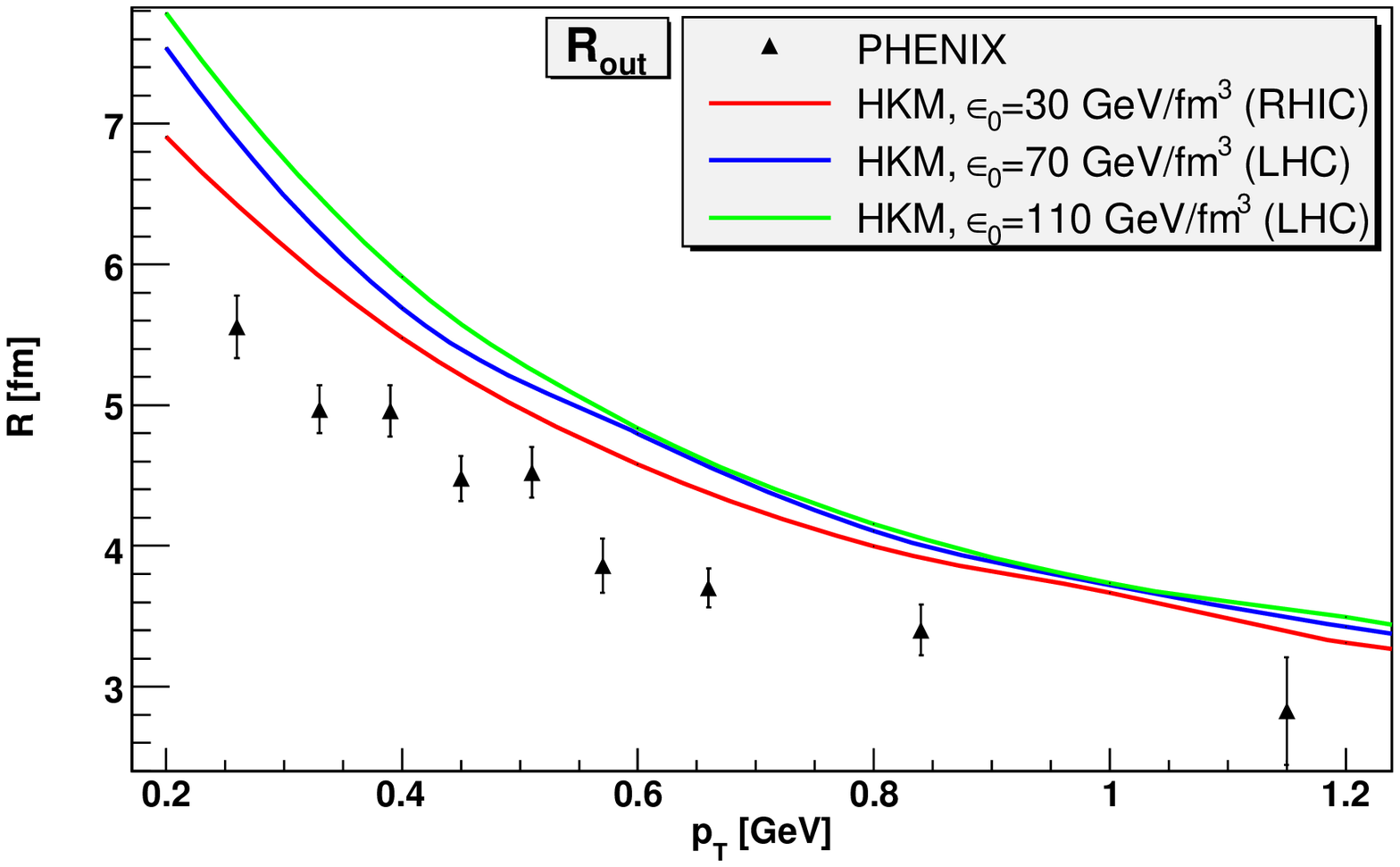}
\end{minipage}

\begin{minipage}[c]{1.0\textwidth}
\hspace{0.0in}
\includegraphics[width=3.2in]{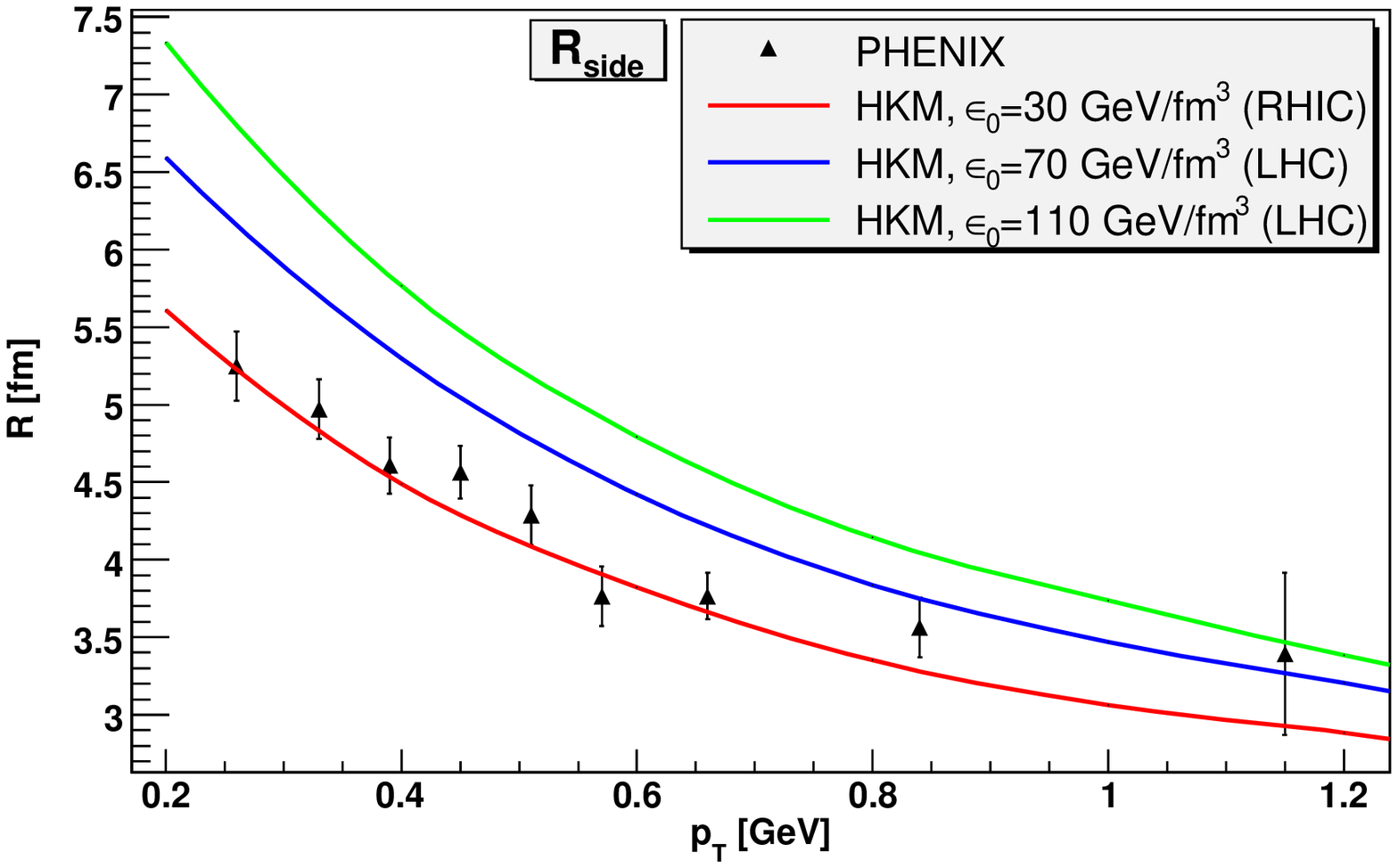}
\hspace{-0.3in}
\includegraphics[width=3.2in]{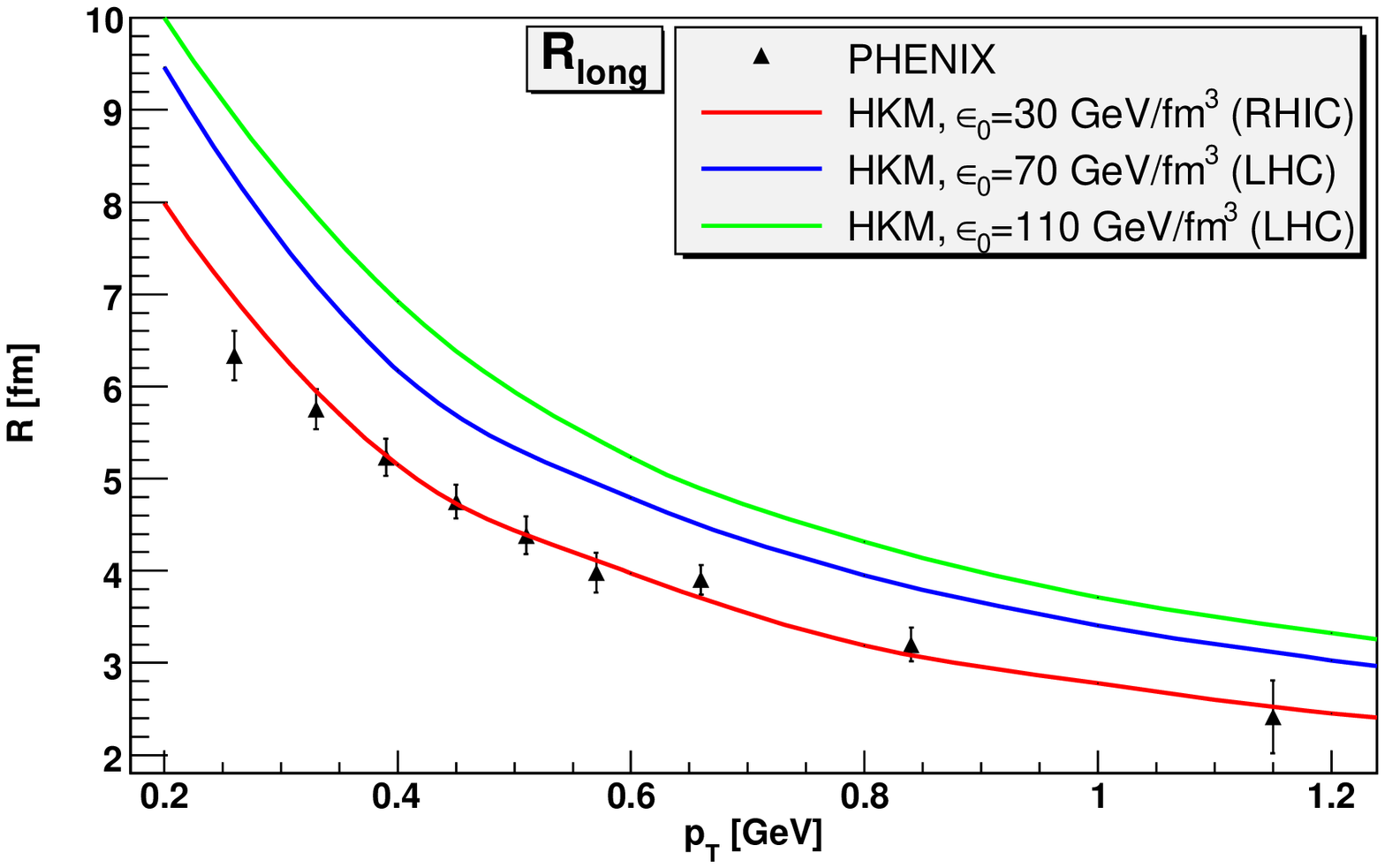}
\vskip -0.3cm
\renewcommand{\captionfont}{\footnotesize \rmfamily}
\renewcommand{\captionlabelfont}{}
\setcaptionwidth{6in}
 \caption{Comparison of the single-particle
momentum spectra of pions  and pion $R_{out}$, $R_{side}$,
$R_{long}$ radii measured by the PHENIX Collaboration for Au+Au
central collisions (HBT radii data were recalculated for $0-5\%$
centrality) at RHIC with the HKM calculations, and HKM predictions
 for Pb+Pb central collisions at LHC. For the sake of
convenience the calculated one-particle spectra  are enhanced in
$1.4$ times.}
\end{minipage}
\label{fig}
\end{narrow}
\vskip -0.8cm
\end{figure}

\end{document}